\documentstyle[draft,psfig]{mn}

\def\kms{\,\hbox{km}\,\hbox{s}^{-1}}
\def\kpc{\,\hbox{kpc}}
\def\Mpc{\,\hbox{Mpc}}

\def\ergscm{\,\hbox{erg}\,\hbox{s}^{-1}\hbox{cm}^{-2}}
\def\keV{\,\hbox{keV}}

\def\ergs{\,\hbox{erg}\,\hbox{s}^{-1}}
\def\Msol{\,M_{\odot}}

\def\lya{Ly-$\alpha$}

\begin{document}

\title[The diffuse Lyman-$\alpha$ halo of LAB1]{Deep SAURON Spectral-Imaging
of the diffuse Lyman-alpha halo LAB1 in SSA22}

\author[R.~G.~Bower et al.]
{R.~G.~Bower$^1$, S.L.~Morris$^1$, R.~Bacon$^2$, R.~J.~Wilman$^1$, 
  M.~Sullivan$^1$, \and S.~Chapman$^3$,
  R.L.~Davies$^4$, P.T.~de Zeeuw$^5$, E.~Emsellem$^2$\\
$^1$Institute for Computational Cosmology, University of Durham, South Road, Durham DH1 3LE, UK\\
$^2$CRAL-Observatoire, 9 Avenue Charles-André, 69230 Saint-Genis-Laval, France\\
$^3$California Institute of Technology, MS 320-47, Pasadena, CA 91125, USA\\
$^4$Department of Physics, University of Oxford, Keeble Road, Oxford OX1 3RH, UK\\
$^5$Sterrewacht Leiden, Postbus 9513, 2300 RA Leiden, the Netherlands}

\maketitle

\begin{abstract}
We have used the SAURON panoramic integral field spectrograph to study the
structure of the \lya\ emission-line halo, LAB1, surrounding the 
sub-millimeter galaxy SMM~J221726+0013. This emission-line halo was 
discovered during a narrow-band imaging survey of the $z=3.1$ large-scale 
structure in the SSA22 region. 
Our observations trace the emission halo out to almost 
$100\kpc$ from the sub-millimeter source and identify two distinct \lya\ 
``mini-haloes'' around the nearby Lyman-break galaxies.
The main emission region has a broad line profile,
with variations in the line profile seeming chaotic and lacking
evidence for a coherent velocity structure. The data also suggests that
\lya\ emission is suppressed around the sub-mm source.
Interpretation of the line structure needs care because \lya\ may be 
resonantly scattered, leading to complex radiative transfer effects, and
we suggest that the suppression in this region arises because
of such effects.  We compare the structure of the central emission-line 
halo with local counter-parts, and find that the emission line halo around
NGC~1275 in the Perseus cluster may be a good local analogue, although the high
redshift halo is  factor of $\sim$100 more luminous and appears to have 
higher velocity broadening. 
Around the Lyman-break galaxy C15, the emission line is narrower, and a
clear shear in the emission wavelength is seen. A plausible explanation 
for the line profile is that the emission gas is expelled from C15 
in a bipolar outflow, similar to that seen in M82.

\end{abstract}

\section{Introduction}

A great deal is now known about the properties of star forming galaxies
in the early universe. Measurements of their clustering properties and 
their luminosity functions has shown that these galaxies are key to
understanding metal enrichment history of the universe (Steidel et al., 1996, 
Adelberger et al., 2003), while spectroscopic studies have established the 
role of galactic ``super-winds'' in regulating the conversion of baryons 
into stars (Pettini et al., 1998, 2001; Teplitz et al., 2000).

One unexpected outcome of emission-line surveys for star forming galaxies
at high redshift has been to establish the
existence of large scale, highly luminous \lya\ haloes, termed `blobs' by
Steidel et al.\ 2000. In this paper, we use the SAURON integral field
spectrograph to study \lya\ blob~1 in SSA22 (hereafter LAB1).
This system is the brighter of two emission haloes that Steidel et al.\ (2000)
discovered during the a survey of the conspicuous spike (at $z=3.07 -
3.11$, Steidel et al., 1998) in the redshift distribution of Lyman-break
galaxies in this field. Subsequently, Chapman et al.\ (2001) discovered
the highly-obscured, very luminous sub-millimeter galaxy (SMM~J221726+0013)
near the centre of this halo. This is possibly a massive
elliptical galaxy seen in formation (Eales et al., 1999, Smail et al., 2002).

Using SAURON, we can map the emission line profiles across the 
LAB1 structure. This allows us to probe the 
nature of the ionised gas surrounding the SCUBA source, gaining
insight into the origin of the diffuse halo (is it primordial
material infalling onto the central object, or material expelled 
during a violent star burst?), the mass of its dark 
matter halo, and the energetics of any super-wind being expelled from
the galaxy. We can also trace the large scale structure
surrounding the central source, and investigate whether similar
haloes surround other galaxies in the field.
The answers to these questions will allow us to understand how galaxy
formation is regulated in massive galaxies in the high-redshift
Universe. They offer key insight into the ``feedback'' process and
will help explain why less than 10\% of the baryon content of the
universe ever forms into stars (the ``cosmic cooling crisis'', 
White \& Rees 1978, Cen \& Ostriker 1999; Balogh et al.\ 2001; Benson et al., 2003).

The gross empirical properties of LAB1 (its spatial extent, luminosity
and proto-cluster environment) are strikingly similar to
those of the extended emission-line regions around luminous $z>2$ radio
galaxies, but there is no powerful radio jet that could be stirring
or ejecting the emitting gas (the radio flux is less than 44$\mu$Jy 
at 1.4 GHz, Chapman et al., 2001; 2003). This is an important distinction: the 
dynamics of the LAB1 halo will give us clearer insight into the confining 
potential and the nature of the emitting material.
Of course, it is likely that the two types of halo are related: Several
authors (eg., Willott et al.\ 2002, Reuland et al.\ 2003) have suggested 
that systems like LAB1 are seen during
an intense star burst phase that is terminated by the triggering of
a powerful radio galaxy.


The layout of the paper is as follows. In \S2, we describe the observations
and the data reduction process. In \S3, we present a quantitative analysis
of the data cube, which we discuss in detail in \S4. We present our 
conclusions in \S5.
Throughout, we assume a flat cosmology with $H_0=70 \kms\Mpc^{-1}$, 
$\Omega=0.3$ and $\Lambda=0.7$. This gives an angular scale at $z=3.1$
of $7.5\kpc/''$.

\section{The Data-Cube}

The SAURON instrument is a high through-put integral field
spectrograph (Bacon et al.\ 2001) that is currently operating on the
William Herschel Telescope, La Palma. It was designed and built by a
partnership between Lyon, Durham and Leiden with the main objective of
studying the dynamics and stellar populations of early-type galaxies
(de Zeeuw et al., 2002).  It combines wide-field
($41^{\prime\prime}\times33^{\prime\prime}$ sampled at
$0.95^{\prime\prime}$) with a relatively high spectral resolution
(4\AA\ FWHM, equivalent to $\sigma=100 \kms$ in the target rest
frame). The instrument achieves this by compromising on the total
wavelength coverage, which is limited to the range from 4810 to
5400\AA. This spatial and spectral sampling ensure that low surface
brightness features are not swamped by read-out noise. However, the
limited spectral coverage means that it is only possible to study the
\lya\ emission from systems at redshifts between $z=2.95$ and
3.45. Fortunately, the SSA22 supercluster lies within this redshift
range. The sky background is devoid of strong night sky emission in
the SAURON wavelength range.  For these observations, the SAURON
grating was upgraded with a volume phase holographic (VPH) 
grism giving an overall system
throughput (including atmosphere, telescope and detector) of
20\%. Because the system uses lenslets (rather than fibres) to
reformat the focal plane, the field of view is fully tiled and no
light is lost between adjacent spectra.

SAURON was used to observe the SSA22 source for a total of 9 hours,
spread over 3 nights in July 2002. The total integration time was built up
from individual 30 min exposures, the telescope being offset by a few
arcsec between each observation. The raw data were reduced using the
XSauron software. The XSauron software is an adaptation for the SAURON
instrument of the public XOASIS software 
(http://www-obs.univ-lyon1.fr/$\tilde{\ }$oasis/home/index.html).
The extraction procedure uses a model
for the instrumental distortions to locate each of the spectra, and
extracts them using optimal weighting, taking account of the 
flux overlap between adjacent spectra. An initial
sky subtraction step was made using SAURON's dedicated sky lenslets that
sample a blank region 2' away from the primary field.
To remove small flat-field and sky subtraction residuals, a super-flat 
was created using the eighteen individual exposures. 
Dividing by this super-flat procedure improved the flat field 
accuracy to 1\% RMS. Each individual datacube was then registered 
to a common spatial location using the faint star in the south east of the
field and then merged into the final data-cube. In the merging process,
we choose to set each spectral-pixel to a size of $1''$ in the spatial 
dimensions and a size of 1.15\AA\ in the wavelength dimension.
In the combined image, the alignment star has a PSF with $1.5''$ FWHM. 
Although weak, large scale residual gradients were evident in the 
combined frame. To produce the map of 
\lya\ emission, we subtracted this residual continuum, using a low order 
polynomial fit to the full wavelength range. This greatly simplifies analysis
of the emission line properties, but means that we cannot attempt to detect
any continuum emission associated with the source of \lya\ photons.
The end result is a 3-D (x,y, $\lambda$) map of the \lya\
emission from the region.

\section{Results}

\begin{figure}
\centerline{\psfig{figure=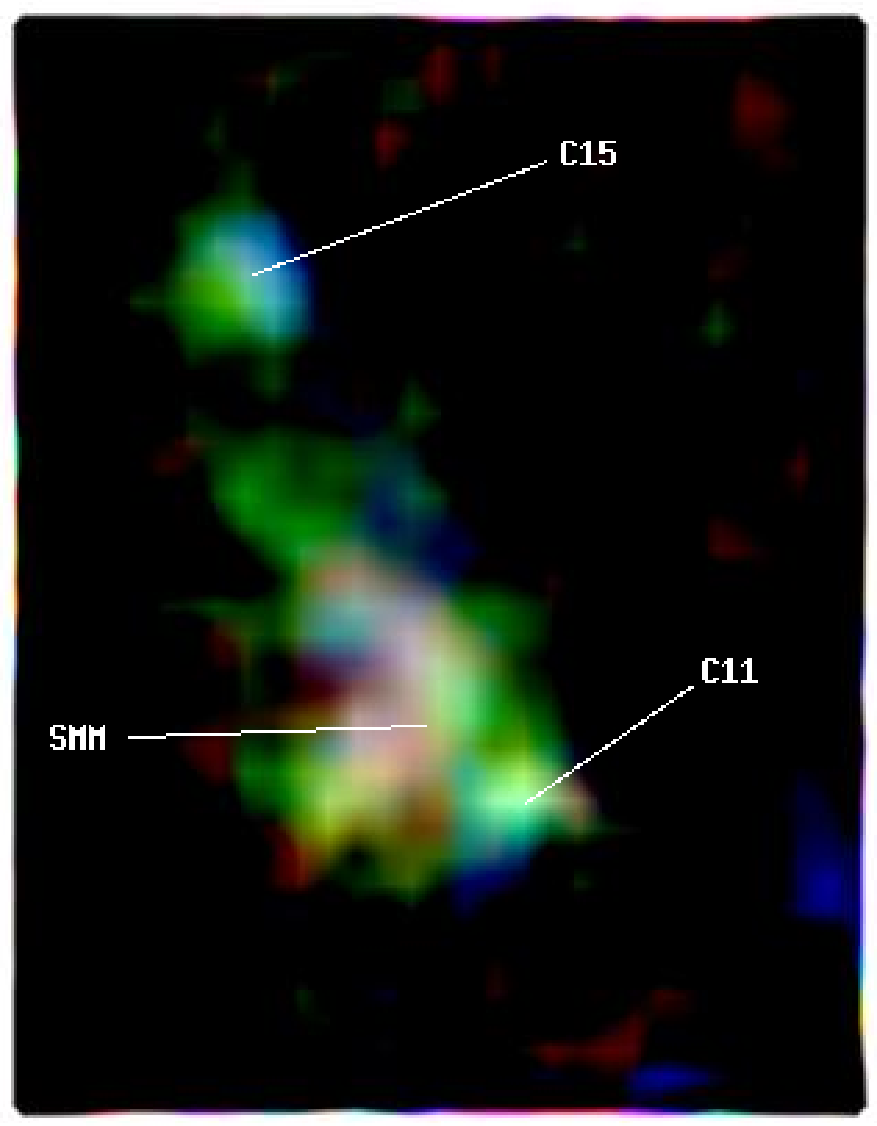,height=4in}}
\caption{A colour representation of the wavelength distribution of \lya\ emission 
in LAB1. A simple interpretation of the image is
that red, green and blue channels represent the red-shifted and blue-shifted 
motions of the ionised material in the halo. The positions of the
two Lyman-break galaxies C11 and C15 are marked, along the position of the
sub-millimeter source (SMM). The area shown is $33'' \times 42''$.}
\label{fig:colour_projection}
\end{figure}

\begin{figure*}
\centerline{\psfig{figure=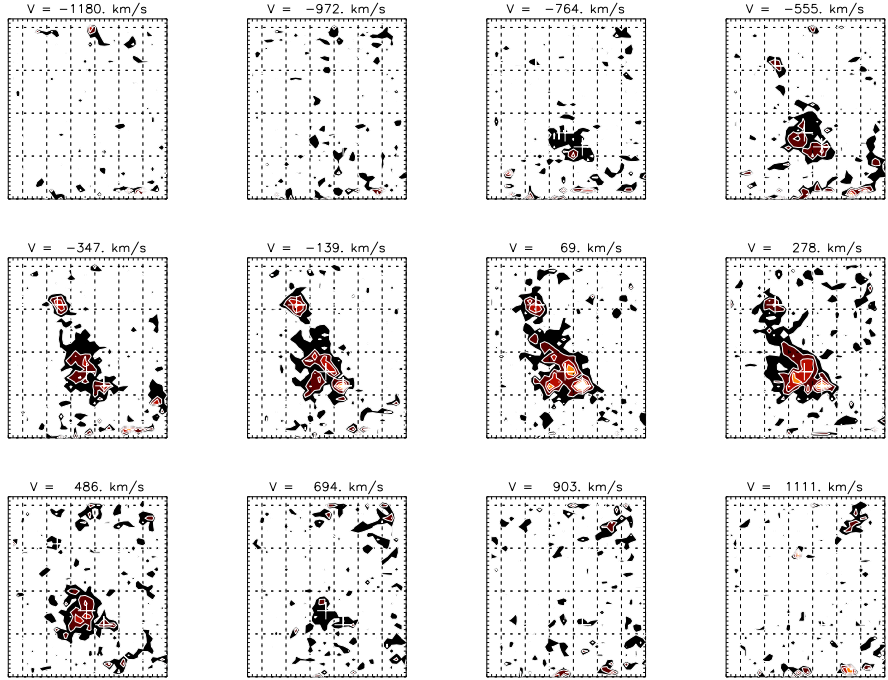,width=5in}}
\caption{A sequence of contour plots showing the changing morphology of 
the \lya\ emission at different wavelengths. The velocity step between
each map is $208 \kms$, with each slice combining a 5.6 \AA\ wavelength 
range so that alternate panels show independent data. Zero velocity 
corresponds to a \lya\ redshift of 3.10. Crosses mark the 
positions of Lyman-break galaxies and the sub-millimeter source. The grid
squares have a spacing of $10''$. The faintest contour has a surface
brightness of $8.6\times10^{-19}\ergscm$ per sq.\ arcsec.}
\label{fig:sequence}
\end{figure*}

The three dimensional data produced by SAURON must be carefully visualised in order to
extract the maximum information from the data. We started by creating a 
colour projection of the data cube shown in Figure~\ref{fig:colour_projection}.
In this view, the red, green and blue colour channels have been constructed from
the data in the wavelength slices centered on 4964.75, 4976.05, 4988.70. 
Each channel in the image is 5.6\AA\ wide ($350 \kms$ in the system rest 
frame).  We have marked the positions
of the Lyman-break galaxies, C11 and C15, identified by Steidel et al.\ 1996, 
2000 and the location of the sub-mm source identified by Chapman et al.\ 2003 
(see below). The data-cube can alternatively be viewed as a sequence
of wavelength slices as shown in Fig.~\ref{fig:sequence}, or these
slices can be combined together to make an animation (this can be viewed at
http://star-www.dur.ac.uk/$\tilde{\ }$rgb/ssa22\_movie.mpg).

Many striking structures can be clearly seen in the main halo. The
overall width of the emission line is very broad ($\sim 500 \kms$ FWHM)
but separate emission structures can be identified. 
There is significant velocity shear in the emission region around the 
Lyman-break galaxies C11 and C15, while the structure 
across the main halo seems more chaotic. The morphology of the diffuse
emission can be seen in these velocity slices: particularly
interesting is the depression seen near the centre of the halo
(this is partially filled by redshifted emission), and the diffuse
extension of the halo towards the near-by Lyman-break galaxy C15. C15
itself is centred in a distinct but much smaller halo. We discuss each of
these features below.

\begin{figure*}
\centerline{\psfig{figure=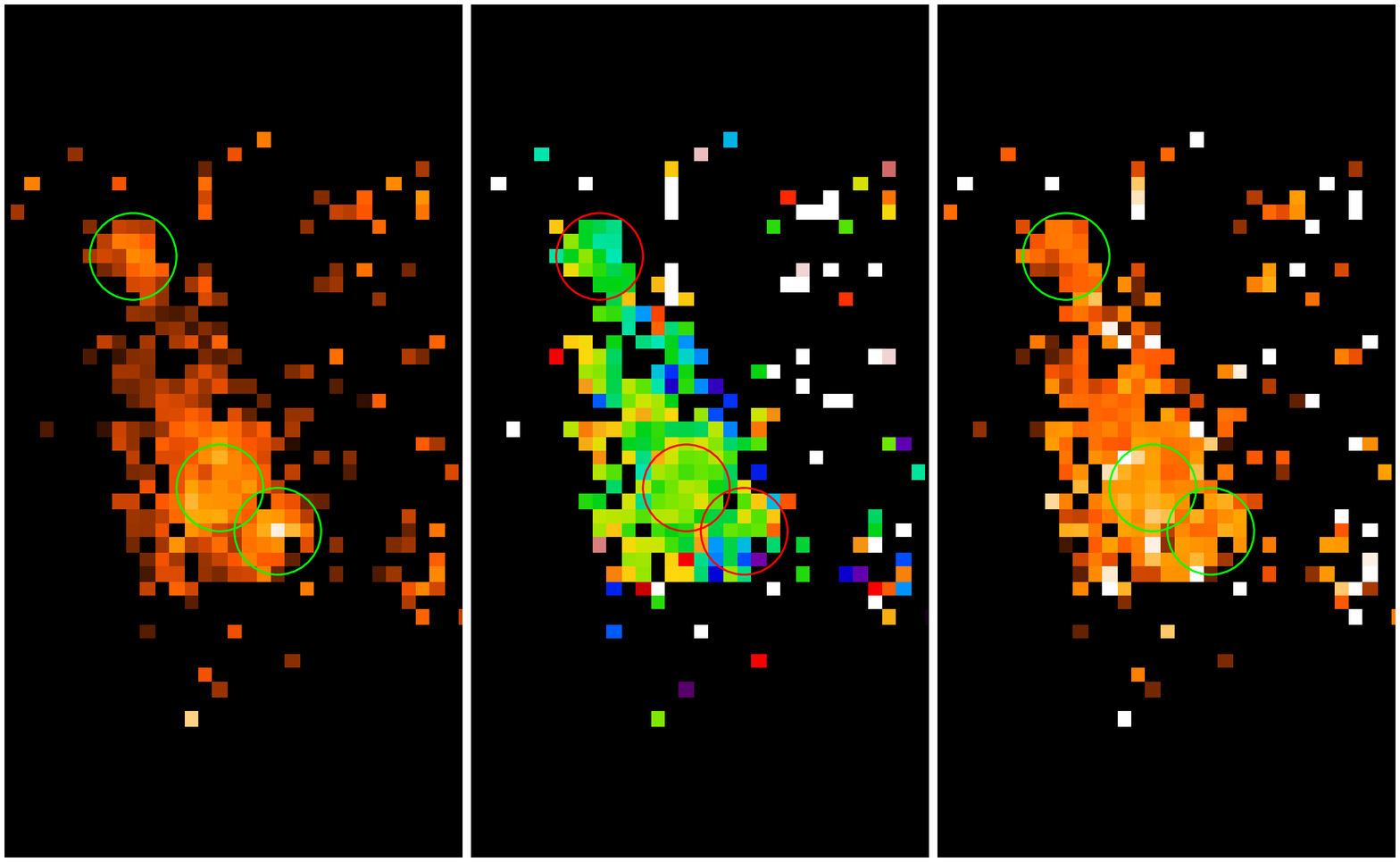,height=3in,angle=270}}
\caption{Single Gaussian fits to the data. Panel (a) shows the
intensity of the fitted line (red--white: 0--$2\times10^{-17}\ergscm$ per
sq.\ arcsec); Panel (b), the central wavelength
of the line (blue--red: 4970--4990\AA); Panel (c), the width of the 
line (red--white: $\sigma = 0$--15\AA).
The plots allow us to quantify the velocity structure seen in the halo.
The circles show the regions used to compute the average line
width and variation in the line centroid. Examples of the line fits
are shown in Figure~\ref{fig:example_fits}.
}
\label{fig:line_fits}
\end{figure*}

\begin{figure*}
\centerline{\psfig{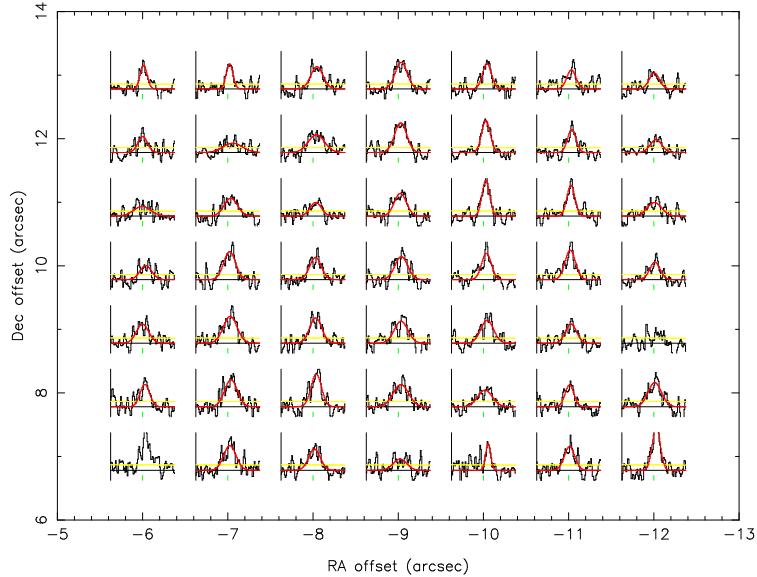}}
\caption{The SAURON spectra (black) and the best fitting Gaussian model
(red), for emission in the region surrounding  SMM~J221726+0013. 
The spectra cover 4930 - 5030\AA\ in the observed frame (equivalent 
to $\pm3000\kms$). The green tick marks 4980\AA. 
The parameters of the line fits are shown in Figure~\ref{fig:line_fits}.
The yellow line shows an estimate of the 1$\sigma$ noise level.
In this co-ordinate system SMM J221726+0013 is roughly centered on
(-9,10) while C11 is centered on (-13,7). 
}
\label{fig:example_fits}
\end{figure*}

To further quantify the emission and its spatial variations, we fitted each
spectrum with a single Gaussian line of variable position, width and
normalisation.  The best fitting parameters for each lenslet
are shown in Fig.~\ref{fig:line_fits}, while some example spectra and 
their associated fits are shown in Fig.~\ref{fig:example_fits}. 
The limiting surface brightness at 
which we are able to reliably detect and fit to the line is $2\times10^{-18} \ergscm$ 
per sq.\ arcsec for lines with $\sigma=5\AA$. Clearly, fitting a single
Gaussian to the emission may not correctly represent the underlying flux
distribution. For this reason, the line fitting approach and the wavelength
slices discussed above should be viewed as complementary rather than 
as alternatives. 

If we concentrate on the emission in a $6''$ diameter circle around the 
central SCUBA source, as shown in the figure, the average velocity width
($\sigma$) is $560\kms$. The shifts in the central wavelength are much smaller than
this: we get an rms lenslet to lenslet variation of only $80\kms$.
Around C11, the line has similar average width ($470\kms$), but the 
rms velocity shift between lenslets is much larger ($240\kms$).
The total flux from the complete region (including C11) that is detected by the line
fitting algorithm is $1.7\times 10^{-15}\ergscm$ (in good agreement with
Steidel et al.\ 2000), corresponding to a 
luminosity of $1.4\times10^{44} \ergs$ in our adopted cosmology.

In the emission region around C15, the line is somewhat narrower ($360\kms$),  
and the rms shift in central velocity ($110\kms$), which
is dominated by the velocity asymmetry that is prominent in the wavelength
slices. The flux from a $6''$ diameter region centered on C15 is 
$1.3\times 10^{-16}\ergscm$.

\begin{figure}
\centerline{\psfig{figure=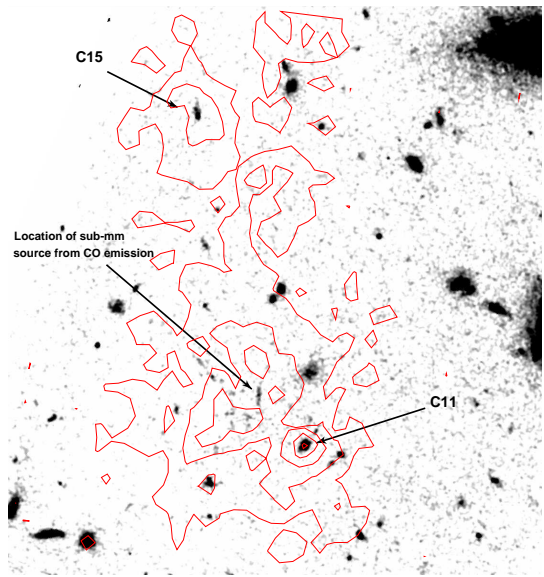,height=3in}}
\caption{A deep STIS image of the SSA22 LAB1 region showing the 
position for the 
SCUBA counterpart (Chapman et al., 2003) relative to the total \lya\ 
emission (contours). The sub-mm source may lie in a 3-D cavity in the emission
(compare contours with Fig.~1). The Lyman-break 
galaxies C15 and C11 are marked: their distinct haloes are clearly seen in 
the 3-D data set.}
\label{fig:overlay}
\end{figure}

The optical counter-part of the sub-mm source has been identified by
Chapman et al.\ (2003; see also Ohyama et al., 2003) after detecting 
the associated CO emission.  To
locate the emission relative to the SCUBA source more precisely, we
aligned the IFU data cube and the HST STIS image of Chapman
et al.\ using the positions of the alignment star and the Lyman
break galaxies C11 and C15. Fig.~\ref{fig:overlay} shows the STIS
image overlayed with the contours of the total \lya\ emission. This
clearly shows the location of the sub-mm source close to the center of
the `cavity' in the emission structure.

\section{Discussion}

Below we divide our results into the separate features seen in our
data and discuss some ideas for how we might interpret them.  The
interpretation is complicated because \lya\ is a resonant
line.  Thus shifts in the feature can appear both because of genuine
gas motion and because photons diffuse in wavelength to escape from
optically thick regions: in regions where the ionised halo is dense
and the gas bulk velocity is relatively constant, 
\lya\ photons must diffuse further into the wings of the line 
to escape. 
We thus expect the changes in the profile to reflect both the bulk 
motion of the material and the variation of optical depth and
dust obscuration with radius (eg., Meinkoln \& Richling 2002). 
In what follows we assume that bulk or turbulent
motion is the dominant source of line broadening. This seems to
be an adequate approximation. Meinkoln \& Richling show that in
static media, the line becomes clearly double peaked before the 
resonant scattering can significantly boost the FWHM line width. 
Their simulations nevertheless show the inherent
difficultly of interpreting the profiles of resonantly scattered 
emission lines. 

\subsection{The velocity structure of LAB1}
The main halo has a complex structure. Within the broad emission,
there are many halo components. The variations in line width and
velocity are inconsistent with a simple outflowing thin shell. The distribution
is better modeled as many distinct gas components, moving relative to each
other with speeds of several hundred $\kms$. One (certainly naive)
interpretation of the wavelength variations is that they reflect the
motions of separate gas clumps bound in a common gravitational potential.

If the above were true, we could use the magnitude of the velocity 
width to infer the halo
mass within $\sim 75 \kpc$ (or $10''$, the typical radius at which the 
emission can be fitted).  If we assume that the clumps are on random 
orbits with a line of sight velocity dispersion of 500 km/s,  the virial theorem 
suggests a mass of order $1.3\times 10^{13} \Msol$, as expected for a 
small cluster with a density distribution given by the Navarro, Frenk \&
White (1997) parametrisation.
It is likely, however, that in fact the clumps have a net outflow or inflow, or
are subject to drag from the intergalactic medium (IGM). This makes the
mass estimate highly uncertain.

It is interesting to compare the \lya\ emission morphology
to the H$\alpha$ emission filaments seen in some local clusters such 
as Perseus. The $H\alpha$ seen around NGC~1275, the central galaxy
of the Perseus cluster, can be traced over a radial extent of $70\kpc$ 
(Conselice, Gallagher \& Wyse, 2001). This is
comparable to the extent of the emission region around SMM~J221726+0013.
Diffuse \lya\ was indeed detected around NGC~1275 with the
IUE satellite (Fabian et al.\ 1984). Fabian et al., estimate a total
\lya\ luminosity of $3 \times 10^{42} \ergs$
based on the assumption that the emission
region has comparable extent to the H$\alpha$ line filaments. This is
a factor 100 smaller than the luminosity of LAB1.
Concelise et al.\ also measure the spread of velocities of the NGC~1275
filaments (over a $14 \kpc$ region). They find a generally chaotic pattern, 
albeit with weak evidence for rotation. The bulk velocity
is comparable to the rms spread in the line centers (both are $ \sim 
150\kms$) but the velocity scale is slower than we find around 
SMM~J221726+0013 (although we recognise that caution is required since we have 
ignored the role of resonant scattering).  Although the luminosity and flow 
velocities around the NGC~1275 system are weaker, it may still be a good local 
analogue for the SMM~J221726+0013 system, and may give us insight into the 
physical origin of the higher redshift halo. Conselice et al.\
conclude that the emission is driven by the interaction of outflowing
material (in the form of hot bubbles rising under their buoyancy) and the 
inflow of material cooling in the cluster potential 
(see also Fabian et al., 2003, Bruggen \& Kaiser, 2003).  

In the high redshift case, cooling alone is unlikely to explain the observed
\lya\ emission, because the \lya\ luminosity is so high compared to the limits
on the X-ray emission ($L_X/L_{\hbox{\lya}} < 10$, Chapman et al., 2003); in a 
intra-cluster medium cooling from a few $\keV$, we would expect 
$L_X/L_{\hbox{\lya}} > 10^3$ even if each hydrogen atom
emits a \lya\ photon (Cowie, Fabian \& Nulsen, 1980). However, at such high redshift
it is far from clear that the gas is ever heated to the virial temperature
of the system. The simulations of Fardal et al.\ 2001 suggest that cooling is so 
rapid that a hydrostatic equilibrium is never established. In their interpretation,
both the extent of the emission halo and the width of the emission line
are due to radiative transfer effects. This seems unlikely since the line-width
appears to increase with radius (rather than decline), 
but much more detailed radiative transfer models are needed to test this 
hypothesis. Our preferred explanation is that the extent and distribution
of line flux reflect the underlying motion and extent of cooling gas.
A powerful collimated outflow appears inconsistent with the lack of velocity shear 
across the halo (unless it is oriented close to the line of sight). The emission 
halo therefore seems best explained by the interaction of slowly rising buoyant material
with cooling gas contained in the proto-cluster potential. The origin of this
hot material remains unclear, however: the convection could equally well be driven by an 
heavily obscured AGN, or by vigorous star formation. 

\subsection{Relation to the sub-mm source}
Figure~\ref{fig:overlay} shows the relative location of the
emission-line halo and the optical counter-part of the strong sub-mm
source (Chapman et al., 2001; 2003). The overlay suggests that the 
sub-mm source may be located at the centre of a `cavity' in the \lya\ 
emission. It has already been established that the \lya\ emission haloes in 
Perseus (Johnstone \& Fabian, 1988) and high redshift radio
galaxies (van Breugel et al., 2003) are ionised by a distributed source.
The presence of the cavity supports this interpretation --- it is 
hard (although not impossible) to construct emission geometries in 
which the central source ionises the surrounding nebulae but is itself 
shielded from the observer.

There are several possible interpretations of the cavity. (1) It may
be a genuine cavity in the ionised gas distribution. This might be
evidence for a strong wind blowing away from the central sub-millimeter
source. However, we would then expect the broadest lines to be observed
close to the center. This is in conflict with the situation observed. 
(2) Figure~\ref{fig:overlay} could also suggest that we are seeing an effect
similar to that responsible for Wide Angle Tail radio sources. The relative
locations of the STIS continuum object compared to the contours of the
Lyman $\alpha$ emission could be consistent with an object blowing out
line emitting gas that is then slowed by ram pressure, losing collimation
in the process and hence producing two large trailing lobes of emission.
The physical details of the emission process are clearly completely
different, but the slowing of the gas by the inter-cluster
medium could be similar. It is unclear, however, why the material would
remain ionised (given its very high luminosity) for a sufficiently long
period of time. 
(3) Alternatively, the cavity may occur because the ionised gas in 
this region contains significant dust (cf., Reuland et al.\ 2003). As
the \lya\ diffuses out of the central region, it is strongly extincted.
This explanation is appealing since we know that the 
central SCUBA source has high extinction. 
If there were no dust at all, we would expect the emission to be scattered
into the line wings rather than being absorbed. This scattering may be 
why the emission that is seen near the location of the sub-mm source 
tends to be redshifted.

\subsection{Mini-haloes around Lyman-break galaxies}
The two other Lyman-break galaxies embedded in the structure
appear to have dynamically distinct haloes. This is particularly clear
for C15 to the north of the main halo (although there is faint emission
that bridges between C15 and the central halo).  A similar halo can also
be discerned around C11. This is a surprising discovery that leads us
to speculate that other Lyman-break galaxies would also have extended
\lya\ haloes if observed to sufficiently low surface brightness.

The mini-halo around the C15 Lyman-break galaxy has its 
own characteristic velocity shear pattern. We can identify the
morphology of this galaxy from the STIS imaging of Chapman et al.\
2003. C15 is elongated at roughly 60 degrees 
(Fig.~\ref{fig:overlay}) to the velocity shear seen in \lya; interestingly,
the morphology of C15 seems somewhat more disturbed than typical Lyman-break 
galaxies. The orientation, together
with the unflattened morphology of the emission,
makes it unlikely that the shear reflects the rotation of a conventional 
gas disk.  Instead, the shear pattern is reminiscent of the
super-wind outflows predicted from proto-galactic disks (Springel \&
Hernquist, 2003), and observed in local starburst galaxies such as M82.  

If we interpret the shear as an outflow inclined at $\sim 45^{\circ}$ to the
line of sight, the pattern suggests a deprojected 
outflow velocity of $\sim 200\kms$, and a physical extent of $\sim 40 \kpc$.
For comparison, the H$\alpha$ emission seen in M82 indicates a deprojected
outflow velocity of $600\kms$ (Shopbell \& Bland-Hawthorn, 1998), extending 
to $11\kpc$ above the disk (Devine \& Bally, 1999). It is interesting that
the dwarf galaxy M82 has a significantly larger outflow velocity than C15 
even though Lyman-break galaxies are thought to have high disk circular 
velocities ($\sim 200 \kms$, Pettini et al., 1998).

However, the geometry of the flow in C15 is qualitatively different,
and does not share the same narrow cone shaped structure as M82 and other local
starburst galaxies (as far as can be discerned from our data). It is 
tempting to suggest that star formation is sufficiently vigorous that the
wind is being driven from across the whole galaxy disk, rather than just
by the nuclear star burst as seen in M82; but examination of
the observed morphology of C15 does not suggest a simple disk galaxy, and 
so a direct morphological comparison is, perhaps, inappropriate.

\section{Conclusions}
 
We have used the SAURON panoramic integral field spectrograph to study the
structure of the \lya\ emission-line halo surrounding the sub-millimetre 
galaxy SMM~J221726+0013. The emission halo can be traced
out to almost $100 \kpc$ from the sub-millimetre source, and the two
nearby Lyman-break galaxies are shown to have kinematically distinct 
emission-line haloes of their own. The main features that we can discern 
are:

\begin{itemize}
\item The emission line profile around the central sub-mm source is broad, 
$\sigma\sim9\AA$. While the line profile varies significantly around 
the sub-mm source, the is no coherent variation in the line centroid.
\item \lya\ emission appears suppressed in the immediate vicinity of the 
sub-mm source.
\item The Ly-break galaxies C15 and C11 appear to be associated with 
enhancements in the emission. These ``mini-haloes'' show significant 
velocity shear. 
\end{itemize}

If we interpret the broad width of the emission line as being due to 
velocity motion of individual gas clouds,
we infer line of sight velocities of $\sim 500 \kms$, suggesting a
dark halo mass of $1.3\times 10^{13} \Msol$ as expected for a small
cluster. We compare the emission
halo to the emission filaments surrounding NGC~1275, the central galaxy
of the Perseus cluster. The chaotic velocity structure and the extent of the
emission are similar, although the \lya\ luminosity of LAB1 is two
orders of magnitude larger. 
Combined with the lack of coherent velocity shear and the high
ratio of the \lya\ and X-ray flux, the comparison leads us to speculate that the 
emission halo of SMM~J221726+0013 is powered by the interaction between 
cooling gas and a relatively weak outflow from the central source.
Our data do not distinguish whether this flow is driven by vigorous star 
formation or by a heavily obscured AGN. It is clear, however, this 
interpretation needs to be confirmed by combining radiative transfer models
with realistic simulations of massive galaxy formation in the
early universe.

The structure of emission halo suggests a cavity around SMM~J221726+0013.
While one possible explanation is that this region has been filled with
hot, completely ionised material, the dip in the emission may equally be 
explained because of dust obscuration in the material ejected from 
the sub-millimeter source. 

The ``mini-haloes'' around the two Lyman-break galaxies in the field
(C11 and C15) show clear velocity shear across their emission haloes. 
The structure
appears to be consistent with a bipolar outflow of material, similar 
to that seen in the star-bursting dwarf galaxy M82. If the material
is an outflow, the deprojected velocity of the flow is $\sim 200\kms$, less 
than the velocity inferred for the outflow from M82, and less than the 
outflow velocities inferred by Pettini et al. (1998, see also Teplitz et al.,
2000 and Pettini et al.\ 2000, 2001) from comparison of the redshifts 
of \lya\ and nebular emission lines in the rest-frame optical. 

These observations break new ground for the SAURON instrument. 
Although it was designed to study the dynamics and stellar populations 
of nearby elliptical galaxies,
we have shown that it can very effectively be used to study low
surface brightness emission features only detectable in long integrations.
It is interesting to speculate how far this powerful technique can be 
taken. On the one hand it is 
key to establish whether the diversity of structure seen in
SSA22 LAB1 is a generic property of other highly luminous sub-mm
galaxies, or whether the deep potential well of the SSA22
super-cluster is necessary to produce emission of this luminosity and
extent. It will also be important to determine whether other Lyman-break
galaxies show mini-halos similar to C15 (cf., Fynbo et al., 2002). 
The observations that we have
presented provide a fore-taste of the deeper observations that will
be made with integral field spectrographs on 8-m telescopes such as GMOS,
(Davies et al.\ 1987), CIRPASS (Parry et al.\ 2000) and SPIFFI 
(Eisenhauer et al.\ 2000, see Bunker et al., 2004 for a recent 
review). Of particular interest is the
combination of large telescope aperture and relatively wide field that
is offered by VIMOS (Le F\`evre et al., 2000) and the planned MUSE 
integral field spectrograph (H\'enault et al., 2003). These instruments
will open the way for deep blank-field surveys of emission-line galaxies 
in the high redshift universe.

\section*{Acknowledgements}

We thank the SAURON instrument team for their support of
this programme, and for creating an instrument with the superb
sensitivity of SAURON. It is a pleasure to thank the ING staff,
in particular Rene Rutten, Tom Gregory and Chris Benn, for
enthusiastic and competent support on La Palma.
The construction of SAURON was made possible through grants
614.13.003 and 781.74.203 from ASTRON/NWO and financial contributions
from the Institut National des Sciences de l'Univers, the Universit\'e
Claude Bernard Lyon-I, the universities of Durham and Leiden, the
British Council, PPARC grant `Extragalactic Astronomy \& Cosmology at
Durham 1998-2002' and the Netherlands Research School for Astronomy NOVA.
RGB is pleased to acknowledge the support of fellowships from PPARC and the 
Leverhulme foundation.


\begin{thebibliography}{}

\bibitem{} Adelberger, K., Steidel, C., Shapley A.E., Pettini M., ApJ, 584 45
\bibitem{} Bacon R., et al., 2001, MNRAS, 326, 23
\bibitem{} Balogh M. L., Pearce F. R., Bower R. G. \& Kay S., 2001, MNRAS, 326, 1228
\bibitem{} Benson A.~J., Bower R.~G., Frenk C.~S., Lacey C.~G., Baugh C.~M., Cole S., 2003, ApJ, 599, 38 
\bibitem{} Bruggen M., Kaiser C., 2003, Nature, 418, 301
\bibitem{} Bunker A. J., Smith J. K., Parry I. R. et al., 2003, Astr.\ Nach., in press (astro-ph/0401002).
\bibitem{} Cen R. \& Ostriker J. P., 1999, ApJ, 514, 1
\bibitem{} Chapman S., Lewis G., Scott D., et al., 2001, ApJ, 548, 17
\bibitem{} Chapman S., Scott D., Windhorst R., Frayer D., Borys C., Lewis G., 2003, in prep.
\bibitem{} Conselice C. J., Gallagher J.S., Wyse R.F.G., 2001, AJ, 122, 2281
\bibitem{} Cowie L. L., Fabian A. C., Nulsen P. E. J., 1980, MNRAS, 191, 399 
\bibitem{} Davies R. L. et al., 1987, SPIE, 2871, 1099
\bibitem{} de Zeeuw T. et al., 2002, MNRAS, 329 513
\bibitem{} Devine D., Bally J., 1999, ApJ, 510, 197
\bibitem{} Eales S. et al., 1999, ApJ, 515, 518
\bibitem{} Eisenhauer F., Tecza M., Mengel S., Thatte N. A., Roehrle C., Bickert K.,  Schreiber J., 2000, SPIE, 4008, 289
\bibitem{} Fabian A. C., Sanders J. S., Crawford C. S., Conselice C. J., Gallagher J. S., Wyse R. F. G., astro-ph/0306039
\bibitem{} Fardal M. A., Katz N., Gardner J. P., Hernquist L., Weinberg D. H., Dav\'e R., 2001, ApJ, 562, 605
\bibitem{} Fynbo J. P. U., Moller P., Thomsen B., et al., 2002, A\&A, 388, 425
\bibitem{} Henault F. et al., 2003, SPIE, 4841, 1096
\bibitem{} Johnstone R. M., Fabian A. C., 1988, MNRAS, 233, 581
\bibitem{} Le Fevre O. et al., 2000, SPIE, 3355, 8
\bibitem{} Meink\"ohn, E., Richling S., 2002, A\&A, 392, 827
\bibitem{} Navarro~J.~F., Frenk~C.~S., White~S.~D.~M., 1997, ApJ, 490, 493
\bibitem{} Ohyama Y., et al., 2003, ApJ, 591, L9
\bibitem{} Parry I. R., Mackay C. D., Johnson R. A., et al., 2000, SPIE, 4008, 1193
\bibitem{} Pettini M., Kellog M., Steidel C. S., Dickinson M., Adelberger K., Giavalisco M., 1998, 508, 539
\bibitem{} Pettini M., Steidel C. S., Adelberger K., Dickinson M.,  Giavalisco M., 2000, ApJ, 528, 96
\bibitem{} Pettini M., Shapley A. E., Steidel C. C., Cuby, J.-G., Dickinson M., Moorwood A. F. M., Adelberger K., Giavalisco M., ApJ, 554, 981
\bibitem{}Reuland M., van Breugel W., Rottgering H., et al., 2003, ApJ, 592, 755
\bibitem{} Shopbell P.L., Bland-Hawthorn J., 1998, ApJ, 493, 129 
\bibitem{} Smail I., Ivison R., Blain W.A., Kneib J.-P., 2002, MNRAS, 331, 495
\bibitem{} Springel V., Hernquist L., 2003, 339, 312
\bibitem{} Steidel, Giavalisco, Dickinson, Adelberger, 1996, ApJ, 462, L17
\bibitem{} Steidel C. C., Adelberger K. L., Dickinson M., Giavalisco M., Pettini M., Kellogg M., 1998, ApJ, 492, 428
\bibitem{} Steidel, Adelberger, Shapley, Pettini, Dickinson, Giavalisco, 2000, ApJ ,532, 170
\bibitem{} Teplitz H.I., McLean I.S., Becklin E.E., 2000, ApJ, 533, L63
\bibitem{} van Breugel W., Reuland M., de Vries W., et al., 2002, astro-ph/0209173
\bibitem[]{} White S.~D.~M., Rees M.~J., 1978, MNRAS, 183, 341 
\bibitem{}Willott C.J., Rawlings S., Archibald E.N., Dunlop J.S., 2002, MNRAS, 331, 435

\end{thebibliography}
\end{document}